\documentclass[twocolumn,
superscriptaddress,
 amsmath,amssymb,
 pre,
]{revtex4-2}

\usepackage{graphicx}
\usepackage{dcolumn}
\usepackage{bm}
\usepackage{color}
\usepackage{siunitx}
\usepackage[dvipsnames]{xcolor}

\definecolor{darkblue}{rgb}{0,0,0.6}
\definecolor{darkred}{rgb}{0.6,0,0}
\usepackage[colorlinks=true,urlcolor=darkblue, citecolor=darkblue, linkcolor=darkred, hyperfootnotes=false]{hyperref}

\newcommand{\ind}[1]{_{\mathrm{#1}}}

\newcommand{\mcA}{\mathcal{A}}
\newcommand{\dd}{\text{d}}
\newcommand{\ed}{\mathrm{e}}

\newcommand{\kk}{\boldsymbol{k}}
\newcommand{\mcL}{\mathcal{L}}

\newcommand{\mcO}{\mathcal{O}}

\newcommand{\mcR}{\mathcal{R}}

\newcommand{\mcV}{\mathcal{V}}

\newcommand{\xx}{\boldsymbol{x}}
\newcommand{\XX}{\boldsymbol{X}}
\newcommand{\yy}{\boldsymbol{y}}

\newcommand{\eeta}{\boldsymbol{\eta}}

\newcommand{\phitw}{\phi^{(2)}}
\newcommand{\phith}{\phi^{(3)}}

\newcommand{\perm}{\mathrm{perm.}}
\newcommand{\transp}{^\mathrm{T}}
\newcommand{\notowner}{\not\owns}

\begin{document}

\title{Algebraic depletion interactions in two-temperature mixtures}

\author{Pascal Damman}
\email{pascal.damman@umons.ac.be}
\affiliation{Laboratoire Interfaces \& Fluides Complexes, Université de Mons, 20 Place du Parc, B-7000 Mons, Belgium}

\author{Vincent Démery}\email{vincent.demery@espci.psl.eu}
\affiliation{Gulliver, CNRS, ESPCI Paris, PSL Research University, 10 rue Vauquelin, 75005 Paris, France}
\affiliation{Univ Lyon, ENS de Lyon, Univ Claude Bernard Lyon 1, CNRS, Laboratoire de Physique, F-69342 Lyon, France}

\author{Guillaume Palumbo}

\affiliation{Laboratoire Interfaces \& Fluides Complexes, Université de Mons, 20 Place du Parc, B-7000 Mons, Belgium}

\author{Quentin Thomas}

\affiliation{Laboratoire Interfaces \& Fluides Complexes, Université de Mons, 20 Place du Parc, B-7000 Mons, Belgium}

\begin{abstract}
The phase separation that occurs in two-temperature mixtures, which are driven out of equilibrium at the local scale, has been thoroughly characterized, but much less is known about the depletion interactions that drive it.
Using numerical simulations in dimension 2, we show that the depletion interactions extend beyond two particle diameters in dilute systems, as expected at equilibrium, and decay algebraically with an exponent $-4$.
Solving for the $N$-particle distribution function in the stationary state, perturbatively in the interaction potential, we show that algebraic correlations with an exponent $-2d$ arise from triplets of particles at different temperatures in spatial dimension $d$.
Finally, simulations allow us to extend our results beyond the perturbative limit.
\end{abstract}

\maketitle

At equilibrium, coupling two objects to a critical field induces an algebraic effective interaction between these objects. The field may be the electromagnetic field, as in the Casimir effect~\cite{Casimir1948,Sushkov2011}, or the density fluctuations in a binary mixture close to its critical point~\cite{Fisher1978,Hertlein2008,Maciolek2018}.
In the latter case, inducing such interactions requires the fine tuning of the temperature~\cite{Maciolek2018}. 
Out of equilibrium, algebraic correlations and the effective interactions that they induce are much more frequent and usually do not require fine tuning.
They arise for instance in driven systems, such as an electrolyte or a binary colloidal mixture under an electric field~\cite{Mahdisoltani2021, Mahdisoltani2021Transient, du2024correlation, Vissers2011Lane, Poncet2017}.
Transient algebraic interactions can also arise during the relaxation following a temperature quench~\cite{Rohwer2017c,Rohwer2018Nonequilibrium}.
Finally, active systems can induce interactions~\cite{Angelani2011}, which may decay algebraically for asymmetric bodies~\cite{Baek2018,Granek2020} or if the active system is flocking~\cite{Fava2024}.
Here, we show that algebraic interactions spontaneously appear in mixtures of particles connected to different thermostats, an out of equilibrium system known to exhibit phase separation~\cite{Awazu2014Segregation, Grosberg2015, Weber2016}.
This algebraic depletion interaction thus does not require self-propulsion and acts between spherical bodies.

Two-temperature mixtures are now a paradigmatic model to study non-equilibrium phenomena.
They are driven out of equilibrium locally, such as active systems, but do not involve self-propulsion; they are thus simpler than the active-passive mixture that they may represent~\cite{McCandlish2012Segregation, Stenhammar2015, Abbaspour2024}.
Two-temperature mixtures can also be mapped to systems with non-reciprocal interactions~\cite{Soto2014Self-Assembly,Ivlev2015Statistical}, which have gained attention recently~\cite{Duan2023Dynamical}.
The hot and cold particles may phase separate even when the interactions between the particles are purely repulsive and identical, provided that the temperature ratio is large enough~\cite{Grosberg2015,Weber2016,Ilker2020}, resulting in dense droplets of cold particles coexisting with a dilute gas of hot particles.
Grosberg and Joanny  provided an analytical prediction for the instability threshold in a dilute system~\cite{Grosberg2015}, showing that denser systems are easier to phase separate. 
Recently, McCarthy \emph{et al} found that the two species may mix again at high density~\cite{McCarthy2024Demixing}.

To understand the phase separation mechanism, it is instructive to consider the depletion interaction between the cold particles that is induced by the hot particles, which is encoded in the pair correlation of the cold particles~\cite{Weber2016}.
In the approach of Grosberg and Joanny, the free energy of the system can be minimized with respect to the density of the hot particles to get an effective free energy functional for the cold particles only~\cite{Grosberg2015}. 
However, this free energy contains only a coarse-grained description of the interaction between the particles, namely the integral of the Mayer function, and does not provide the shape of the effective interaction.
An explicit expression of the effective interaction has been obtained in Ref.~\cite{Ilker2020} using the potential of mean force; however, the application of this approach to an out-of-equilibrium system is questionable.
Indeed, we will see that generalizing different equilibrium methods to compute the effective interactions in this out-of-equilibrium system leads to different results, similarly to the situation observed with the pressure in a gas of active particles~\cite{Solon2015Pressure}.

In this Letter, we investigate the depletion interaction between cold particles that is due to the hot particles in the dilute limit using analytical calculations and numerical simulations.
Restricting ourselves to the dilute limit amounts to consider a system of two cold particles and a single hot one.
Contrary to two-body problems~\cite{Grosberg2015, Soto2014Self-Assembly}, this three-body system cannot be mapped onto an effective equilibrium system~\cite{Wang2020Three-body}.
We compute the depletion interaction analytically, perturbatively in the interaction strength $\epsilon$.
At order $\epsilon^2$, we show that the potential of mean force argument should be corrected by a purely non-equilibrium three-body potential.
At order $\epsilon^3$, we find that the three-body potential gives rise to algebraic interactions decaying as $r^{-2d}$ in spatial dimension $d$.
Simulations confirm our analytical results for weak interactions and show that the algebraic decay is still present for strong interactions, allowing us to discuss the dependence of the magnitude of this decay on the interaction potential.

We consider particles with positions $\xx_i(t)$, connected to thermostats with temperature $T_i$ and obeying an overdamped Langevin dynamics:
\begin{equation}\label{eq:model}
  \dot\xx_i(t)=-\sum_j\nabla V(\xx_i(t)-\xx_j(t))+\sqrt{2T_i}\eeta_i(t),
\end{equation}
where $V(r)=\epsilon U(r)$ is the interaction potential.
The vectorial Gaussian white noises $\eeta_i$ have a unit variance and are uncorrelated.
The particles mobility and the Boltzmann constant have been set to unity.
We consider only two species $\alpha\in\{A,B\}$ connected to thermostats $T_\alpha$.
To focus on the interactions induced by the particles $B$ on the particles $A$, we assume that there are only two $A$ particles $i\in\{1,2\}$ and a small density $\rho$ of $B$ particles. 
We expand the pair distribution function of the two $A$ particles $g(r)$ as $g(r)=\exp(-V(r)/T_A)\left[1+\sum_{n=1}^\infty \rho^n g_n(r)\right]$~\cite{Hansen2006}.
We focus on the first term of the expansion, $g_1(r)$; we denote it $\hat g(r)$ and refer to it as the ``correlation'':
\begin{equation}\label{eq:rcd}
\hat g(r)= \lim_{\rho\to 0}\frac{e^{V(r)/T_A}g(r)-1}{\rho}.
\end{equation}
This order is not affected by the interactions between the $B$ particles, which can render the depletion interaction nonmonotonic~\cite{Mao1995Depletion}.

At equilibrium, here when $T_A=T_B=T$, the interaction induced by the $B$ particles can be obtained by integrating the three body distribution $f(\xx_1, \xx_2, \xx_3)$ over the coordinate of the third particle, leading to~(App.~\ref{sec:eq_distrib})
\begin{equation}\label{eq:eq}
  \hat g_{\mathrm{eq}}(r) =  \left(e^{-V/T}-1 \right)*\left(e^{-V/T}-1 \right)(r),
\end{equation}
where the star denotes the convolution product.
In the weak interaction limit $\epsilon\ll T$ it reduces to
\begin{equation}\label{eq:eq_weak}
  \hat g_{\mathrm{eq}} (r)= \frac{V*V(r)}{T^2}  + \mcO(\epsilon^3).
\end{equation}
If the potential $U(r)$ has a finite range $\sigma$, which represents the diameter of the particles, both the full expression~(\ref{eq:eq}) and its weak interaction limit~(\ref{eq:eq_weak}) are zero beyond two particle diameters.
The same expressions can be recovered using the potential of mean force~\cite{Hansen2006}~(App.~\ref{sec:eq_pmf}).

These arguments can be transposed out of equilibrium, using the fact that an isolated pair of particles $(i,j)$ is in an effective equilibrium at temperature $T_{ij}=(T_i+T_j)/2$ so that its pair correlation is $g_{ij}(r)=\exp(-v_{ij}(r))$, where $v_{ij}(r)=V(r)/T_{ij}$~\cite{Grosberg2015}.
Using the potential of mean force, Ilker and Joanny obtained~\cite{Ilker2020} (App.~\ref{sec:noneq_pmf})
\begin{align}
\hat g(r) &=  \frac{T_B}{T_A}\left(e^{-V/T_{AB}}-1 \right)*\left(e^{-V/T_{AB}}-1 \right)(r) \label{eq:pmf}\\
& = \frac{T_B}{T_A T_{AB}^2} V*V(r) + \mcO(\epsilon^3). \label{eq:pmf_weak}
\end{align}
Alternatively, assuming that the $N$-particle distribution function is given by $f(\XX)=\exp \left( -\sum_{\langle ij\rangle} v_{ij}(\xx_i-\xx_j) \right)$, where $\XX=(\xx_i)_{1\leq i\leq N}$, we get~(App.~\ref{sec:noneq_distrib})
\begin{align}
  \hat g(r) &=\left(e^{-V/T_{AB}}-1 \right)*\left(e^{-V/T_{AB}}-1 \right)(r) \label{eq:from_pair}\\
  & = \frac{V*V(r)}{T_{AB}^2} + \mcO(\epsilon^3). \label{eq:from_pair_weak}
\end{align}
The temperatures enter differently into the predictions~(\ref{eq:pmf}) and~(\ref{eq:from_pair}), pointing to the fact that these approaches may not apply out of equilibrium.
However, they agree on the form of the depletion interaction, which is the same as the equilibrium one; in particular its range is limited to two particle diameters.

\begin{figure*}
    \begin{center}
        \includegraphics[width=.32\linewidth]{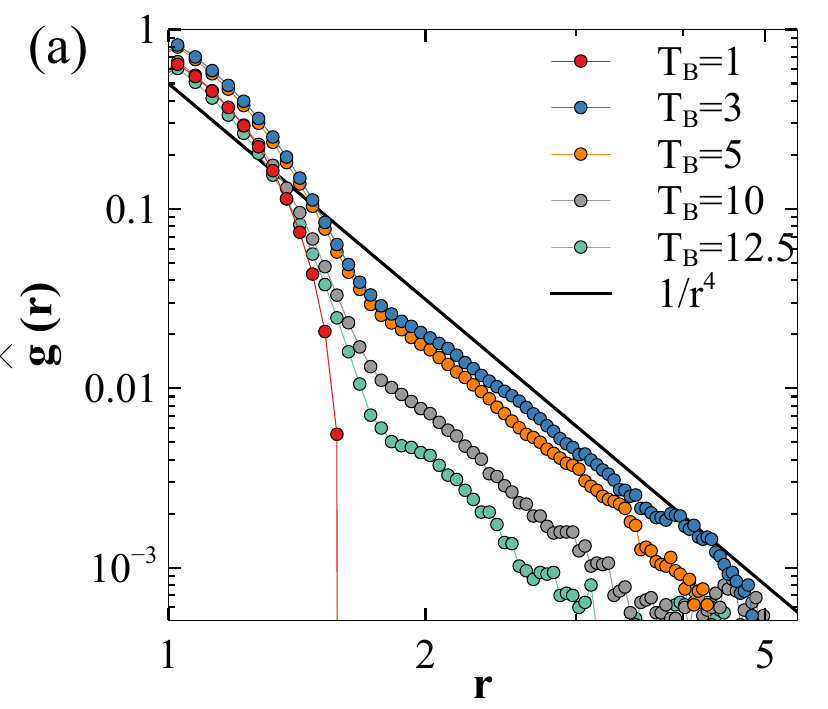}
        \includegraphics[width=.32\linewidth]{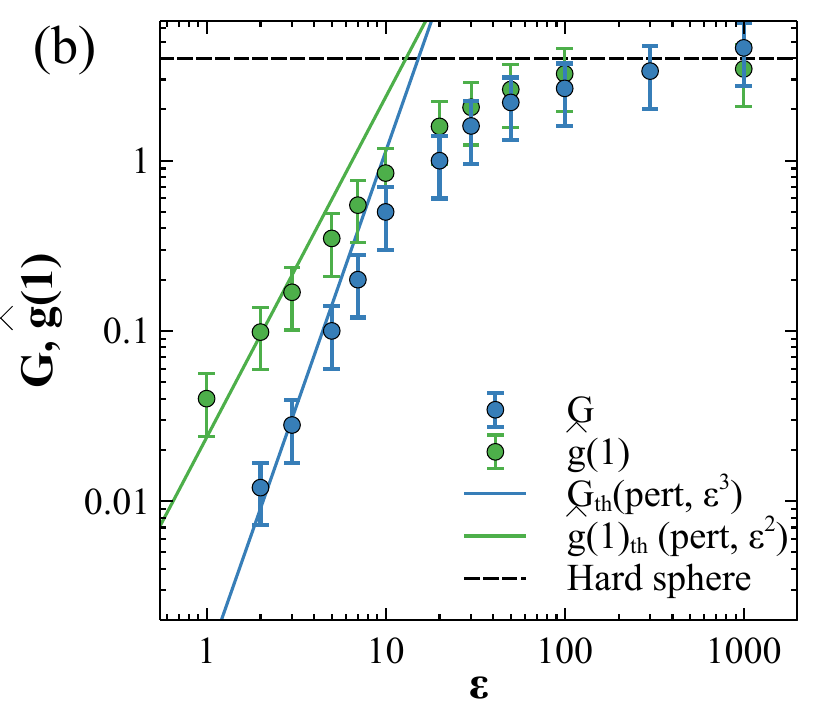}
        \includegraphics[width=.32\linewidth]{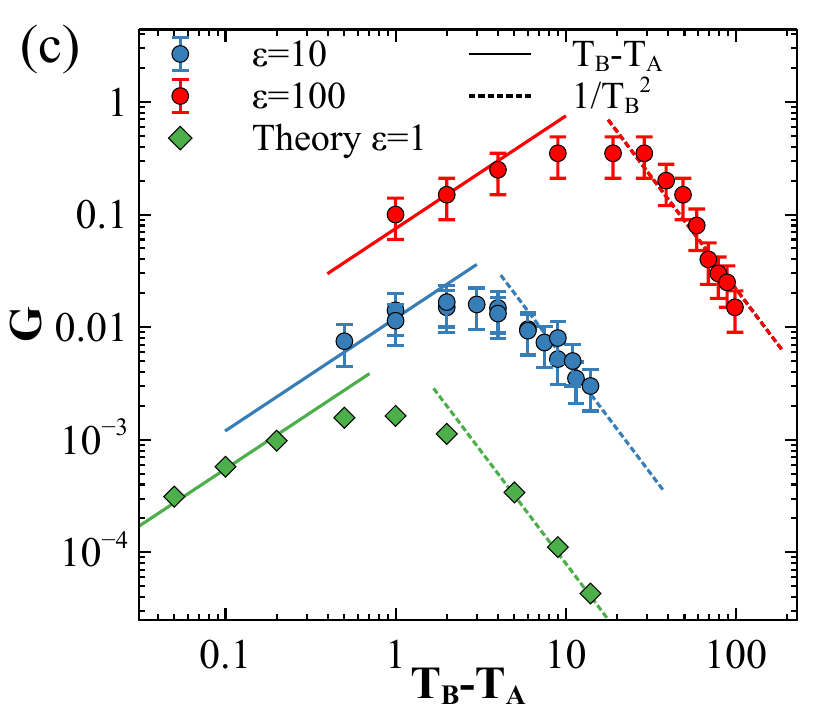}
    \end{center}
    \caption{(a) Correlation function $\hat{g}(r)$ of the species $A$ for $T_A=1$, $\epsilon=10$ and $\rho=0.05$ ($T_B$ as indicated).  
  (b) Correlation at contact $\hat g(1)$ and prefactor of the algebraic decay $G$ as a function of $\epsilon$ ($T_A=1, T_B=3$). The solid lines correspond to the theoretical predictions.
    (c) Prefactor of the algebraic decay $G$ as a function of $T_B-T_A$ and theoretical prediction ($\epsilon$ as indicated). 
    \label{fig:distrib}}
\end{figure*}

We simulated Eq.~(\ref{eq:model}) numerically~\cite{LAMMPS} (App.~\ref{sec:details_numerics}) in spatial dimension $d=2$ for three different interactions between particles with diameter $\sigma=1$: 
harmonic, $U\ind{harm}(r)=(1-r)^2\theta(1-r)$, $\theta$ being the Heaviside function, 
Gaussian, $U\ind{Gauss}(r)=\exp(-6r^2)$, 
and WCA, $U\ind{WCA}(r)=(r^{-6}-1)^2\theta(1-r)$.
In order to improve the statistics, we used the same small density $\rho$ of $A$ and $B$ particles.
In this situation, an additional depletion interaction is induced by the particles $A$ themselves, so that $\hat g(r)=\hat g^A(r)+\hat g^B(r)$, where $\hat g^A(r)=\hat g_{\mathrm{eq}}(r)$ (Eq.~(\ref{eq:eq})).
The correlations $\hat g(r)$ calculated from simulations with an harmonic interaction, $\epsilon=10$, $\rho=0.05$, $T_A=1$ and different values of $T_B$ are presented in Fig.~\ref{fig:distrib}(a).
As expected, we observe an increase of the correlation at contact $\hat g(1)$ due to the depletion interaction.
This increase scales as $\epsilon^2$ with the interaction strength $\epsilon$, as expected from the different theoretical predictions (Fig.~\ref{fig:distrib}(b)).
However, in contradiction with the predictions, the depletion interaction extends beyond two diameters as soon as the system departs from equilibrium.
Beyond two particle diameters, the radial dependence of the depletion interaction is compatible with an algebraic decay, $\hat g(r)=G/r^4$.
The prefactor $G$ of the algebraic decay increases with the interaction strength as $\epsilon^3$ for small $\epsilon$ and saturates for large values of $\epsilon$, corresponding to the hard sphere limit (Fig.~\ref{fig:distrib}(b)).
The effect of the temperature $T_B$ is non-monotonic: $G$ first increases and then decreases at large temperatures (Fig.~\ref{fig:distrib}(c)).
The temperature where the effect is maximal increases with the interaction strength $\epsilon$.
We now return to theory to (i) elucidate the discrepancy between the two arguments adapted out of equilibrium and (ii) find the origin and the characteristics of the algebraic decay.

We use two theoretical approaches.
First, we use Stochastic Density Field Theory~\cite{Dean1996} to turn the microscopic dynamics~(\ref{eq:model}) into exact Langevin equations for the density fields $\hat\rho_\alpha(\xx,t)=\sum_{i\in I_\alpha}\delta(\xx-\xx_i(t))$, where $\alpha\in\{A,B\}$ indicate the species and $I_\alpha$ is the set of indices of the particles of species $\alpha$.
SDFT applies in equilibrium as well as in out-of-equilibrium situations~\cite{Demery2016Conductivity, Poncet2017, Benois2023Enhanced}.
The exact dynamics of the density fields is non-linear due to pair interactions, and contains multiplicative noise, which makes it intractable in practical situations. 
In the limit of a dense system with weak interactions, SDFT can be linearized and the density fluctuations around the average densities $\rho_\alpha$ become Gaussian~\cite{Demery2014c,Dean2014}; this approximation corresponds to the random phase approximation in liquid theory~\cite{Hansen2006}.
The correlation, which corresponds to the correlations of the fluctuations of the density fields, can be calculated in Fourier space~(App.~\ref{sec:sdft}).
In the dilute limit that we consider here ($\rho_A=0$, $\rho_B\to 0$), it reduces to
\begin{equation}\label{eq:sdft_dilute}
    \hat g(r)=\frac{V*V(r)}{T_A T_{AB}}+ \mcO(\epsilon^3).
\end{equation}
This expression is in quantitative agreement with the simulations for $r\lesssim 2$ (App.~\ref{sec:num_int}, Fig.~\ref{fig:quantitative_comp}).
It takes a similar form as the results obtained by adapting equilibrium arguments (Eqs.~(\ref{eq:pmf_weak}, \ref{eq:from_pair_weak})), but the temperatures enter differently in the prefactor.
As the expression~\eqref{eq:sdft_dilute} is exact at the order $\epsilon^2$, we conclude that none of the two equilibrium calculations can be adapted out of equilibrium.

\begin{figure}
    \begin{center}
           \includegraphics[width=.8\linewidth]{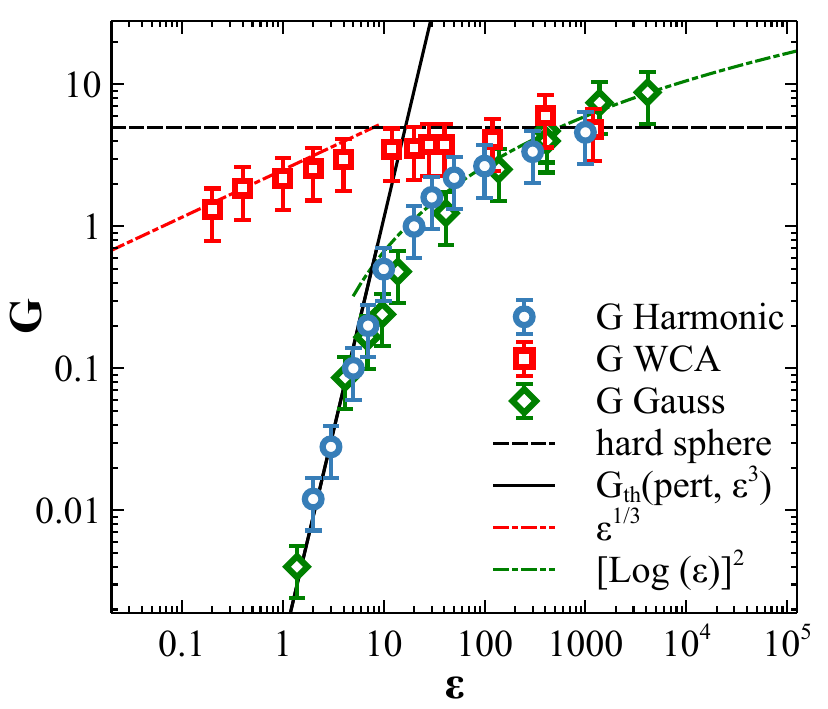}
    \end{center}
    \caption{Prefactor $G$ of the algebraic decay as a function of the interaction strength $\epsilon$ for the harmonic (blue circles), Gaussian (green diamonds), and WCA (red squares) interactions from the numerical simulations.
    The lines represent the several asymptotic expressions: weak interaction for an integrable potential (black solid line), hard spheres (black dashed line), weak interaction for a WCA potential (red dash-dotted line) and strong interaction from a Gaussian interaction (green dash-dotted line).
    \label{fig:prefactor}}
\end{figure}

Yet, the expression (\ref{eq:sdft_dilute}) vanishes at two particle diameters and does not explain the algebraic decay observed in the simulations (Fig.~\ref{fig:distrib}(a)).
This is expected, because the simulations indicate that the algebraic decay arises at order $\epsilon^3$, while Eq.~(\ref{eq:sdft_dilute}) is limited to order $\epsilon^2$.
We now turn to a small density expansion that is valid for any interaction strength.
The $N$-particle distribution $f(\XX)$ is stationary under the Smoluchowski equation describing the microscopic dynamics~\eqref{eq:model}:
\begin{equation}
 \partial_t f = \sum_i\nabla_i\cdot \left[T_i\nabla_i f(\XX) + f(\XX)\sum_{j\neq i}\nabla V_{ij}(\xx_i-\xx_j) \right].
\end{equation}
To isolate the many-body effects from the pair correlation obtained by Grosberg and Joanny~\cite{Grosberg2015}, we write it as
\begin{equation}
f(\XX)=\exp\left(-\sum_{\langle ij\rangle}v_{ij}-\phi(\XX)\right),  
\end{equation} 
where now $v_{ij}=V(\xx_i-\xx_j)/T_{ij}$.
The resulting equation for the many-body potential $\phi(\XX)$ is~(App.~\ref{sec:mbp})
\begin{multline}\label{eq:phi}
0=\sum_iT_i\left[(\nabla_i\phi)^2-\nabla_i^2\phi\right]\\+\sum_{\langle ij\rangle}\nabla v_{ij}\cdot[(T_{ij}+\tau_{ij})\nabla_i\phi-(T_{ij}-\tau_{ij})\nabla_j\phi]\\
+\sum_{\langle ijk\rangle}\left(\tau_{ijk}\nabla v_{ij}\cdot\nabla v_{ik}+\perm\right),
\end{multline}
where we have defined the temperature differences $\tau_{ij}=T_i-T_j$ and $\tau_{ijk}=T_i-T_{jk}$ and ``$\mathrm{perm}$'' indicates the two terms in the last sum obtained from the first by cyclic permutations.
The third term is the source term: it is a sum over triplets, underlining the fact that $\phi$ originate from three-body effects. 
Moreover, it involves temperature differences so that it is zero at equilibrium.
While this is expected, it stresses the fact that the $N$-particle distribution can be written as a product of pair terms only at equilibrium, and that three-body effects arise as soon as the system is put out of equilibrium, as evidenced for particles attached by linear springs~\cite{Wang2020Three-body}.
The equation~\eqref{eq:phi} for $\phi$ is a nonlinear PDE with multiplicative terms, and we solve it perturbatively in the interaction strength $\epsilon$: we write $\phi=\sum_{k=2}^\infty \phi^{(k)}$, with $ \phi^{(k)}\propto\epsilon^k$.

The depletion interaction encoded in $\hat g(r)$ can be computed from the three-body distribution $f(\xx_1, \xx_2, \xx_3)$.
Expanding $f$ in powers of $\epsilon$ and integrating over $\xx_3$, we find~(App.~\ref{sec:correc_correl_gen})
\begin{multline}
    \hat g=\int \dd\xx_3\left[v_{13}v_{23}-\phi^{(2)}-\frac{1}{2}\left(v_{13}^2v_{23}+v_{13}v_{23}^2\right)\right.\\
    \left.+(v_{13}+v_{23})\phi^{(2)}-\phi^{(3)}\right] + \mcO(\epsilon^4).
    \label{eq:y_dil}
\end{multline}
The first two terms in the integrand are of order $\epsilon^2$, while the last three are of order $\epsilon^3$.

At order $\epsilon^2$, the Laplacian in the first term balances the source term, and the solution is given by $\phitw(\XX)=\sum_{\langle ijk\rangle}\tau_{ijk}\omega^{(2)}_{ijk}(\xx_j-\xx_i, \xx_k-\xx_i)+\perm$, where $\omega^{(2)}_{ijk}$ is given in Fourier space by~(App.~\ref{sec:mbp2})
\begin{equation}
    \tilde\omega^{(2)}_{ijk}(\kk,\kk') = \frac{\kk\cdot\kk'\tilde v_{ij}(\kk)\tilde v_{ik}(\kk')}{2(T_{ij}k^2+T_{ik}k'^2+T_i\kk\cdot\kk')}.
\end{equation}
Using $\phitw$ to compute $\hat g(r)$ at order $\epsilon^2$, the expression~\eqref{eq:sdft_dilute} obtained from SDFT is recovered~(App.~\ref{sec:correc_correl_2}).
This result shows that the three-body effects are important to get the correct depletion interaction at order $\epsilon^2$, and that these effects are included in the SDFT calculation.

To find $\phith$, we keep the terms of order $\epsilon^3$ in Eq.~\eqref{eq:phi}:
\begin{multline}\label{eq:phi33}
\sum_i T_i \nabla_i^2\phi^{(3)} \\ =  \sum_{\langle ij \rangle} \nabla v_{ij}\cdot \left[(T_{ij}+\tau_{ij})\nabla_i\phi^{(2)} - (T_{ij}-\tau_{ij})\nabla_j\phi^{(2)} \right],
\end{multline}
which can be solved using the expression of $\phitw$ obtained previously~(App.~\ref{sec:mbp3}).
Using $\phitw$ and $\phith$ in the correction~\eqref{eq:y_dil} leads the depletion interaction at order $\epsilon^3$.
The resulting expression contains many terms involving double integrals in Fourier space, which can be performed numerically, yielding a good agreement with the $\hat{g}(r)$ obtained from simulations~(App.~\ref{sec:correc_correl_3}, Fig.~\ref{fig:quantitative_comp}).
The small wavevector behavior in Fourier space leads to an algebraic decay:
\begin{equation}\label{eq:gen_conj}
  \hat g(r) \underset{r\to\infty}{\sim} G_U\left(\frac{\epsilon}{T_A}, \frac{T_B}{T_A} \right)r^{-2d},
\end{equation}
where
\begin{equation}
  G_U(e,\theta)\underset{e\to 0}{\sim}\left[e\tilde U(0)\right]^3F(\theta). 
\end{equation}
This prediction agrees quantitatively with the simulations for the harmonic and Gaussian interactions, which are integrable (Fig.~\ref{fig:prefactor}).
The temperature dependence follows $F(\theta)\sim\theta-1$ close to equilibrium ($\theta\to 1$), consistent with the fact that the algebraic interaction emerges from the temperature differences. 
When the $B$ particles are very hot ($\theta\to \infty$), they are barely affected by the interactions and the prefactor decays as $F(\theta)\sim\theta^{-2}$.
These behaviors are confirmed by the simulations (Fig.~\ref{fig:distrib}(c)).

Furthermore, the simulations show that the algebraic decay as $r^{-2d}$ holds for the three different interactions, harmonic, WCA, Gaussian, independently of their strength (App.~\ref{sec:sim_gauss_wca}, Fig.~\ref{fig:size}(a,b)).
Hard spheres are obtained as the strong interaction limit for interactions with a finite range $\sigma$.
In this limit, the prefactor depends only on the diameter $\sigma$ (Fig.~\ref{fig:prefactor}).
Assuming that the correlation depends only on the dimensionless ratio $r/\sigma$, the prefactor should thus be 
\begin{equation}
  G_U(e, \theta)\underset{e\to\infty}{\to} \sigma^{2d}F\ind{hs}(\theta);
\end{equation}
this argument is confirmed by simulations (Fig.~\ref{fig:size}(c)).
Two situations remain uncovered by the analysis above: the WCA interaction in the soft limit, because it is not integrable, and the Gaussian interaction in the hard limit, because its range is not finite.
They can be adressed by defining the effective diameter $\sigma\ind{eff}$ as $\epsilon U(\sigma\ind{eff})=T_A$, and then using it as an effective hard sphere diameter.
For the WCA interaction in the limit $\epsilon\to 0$, this gives $\sigma\ind{eff}^\mathrm{WCA}\sim (\epsilon/T_A)^{1/12}$ and $G_U(e, \theta)\sim e^{1/3}$.
For the Gaussian interaction in the limit $\epsilon\to\infty$, this gives $\sigma\ind{eff}^\mathrm{Gauss}\sim \sqrt{\log(\epsilon/T_A)}$ and $G_U(e,\theta)\sim [\log(e)]^{2}$.
These behaviors are compatible with the simulations (Fig.~\ref{fig:prefactor}).
Finally, we ran numerical simulations with a Lennard-Jones interaction, which has an attractive part.
Two behaviors are observed: if the attraction is small enough, the system remains homogeneous and the algebraic decay of the correlation is observed; if the attraction is too strong, the cold particles aggregate and the algebraic decay is lost~(App.~\ref{sec:sim_lj}, Fig.~\ref{fig:hat_g_LJ}).

The depletion interaction that we have unveiled is the driving force behind the formation of dense droplets of cold particles when they separate from the hot particles.
As the algebraic tail of the interaction originates from temperature differences ($\tau_{ijk}$ in Eq.~(\ref{eq:phi})), this effect is reversed if one considers the depletion interaction induced by the cold particles on the hot ones, so that the algebraic tail is \emph{repulsive} in this case. 
This repulsion between hot particles may work with the attraction between cold particles to trigger the phase separation.
Our results may also apply to mixtures of passive and artificial~\cite{Singh2017, Madden2022Hydrodynamically} or living~\cite{Ortlieb2019, Lagarde2020} self-propelled particles, which could be described at large scales by two-temperature mixtures.

\begin{acknowledgements}
The authors acknowledge fruitful discussions with Denis Dumont and Cesare Nardini and the support by F.R.S.-FNRS under the research Grant (PDR ''Active Matter in Harmonic Trap'') No. T.0251.20.
This work was supported by the Belgian National Fund for Scientific Research (FRS-FNRS) within the Consortium des \'equipements de Calcul Intensif – CECI, under Grant 2.5020.11t. Quentin Thomas is FRIA grant holder from the Belgian National Fund for Scientific Research (FRS-FNRS).
\end{acknowledgements}

\appendix

\section{Details of the numerical simulations}
\label{sec:details_numerics}

The simulations were performed with the LAMMPS package~\cite{LAMMPS}. The systems consist in $N=\num{20000}$ particles with $N/2$ thermalized at $T_A$ and $N/2$ thermalized at $T_B$. 
We follow the Brownian Dynamics time integration to update the positions.
This integration scheme solves a system of $N$ overdamped Langevin equations,
\begin{equation}
\dd r_i = \gamma_i^{-1} F_i \dd t + \sqrt{2k_B T_i} \gamma_i^{-1/2} \dd W_{i,t}
\end{equation}
where all friction coefficients $\gamma_i$ are set to unity. The $\dd W_{i,t}$ represent Wiener processes mimicked by discrete sequences of Gaussian random numbers with zero mean and unit variance.

The timestep was adjusted to ensure the stability and reproducibility of the simulations.
The interaction between all particles were described with the pair style command and the potentials given in the manuscript.
Finally, to improve the accuracy of the data, the correlation $\hat g(r)$ has been averaged on a large number of simulations ($\numrange{1000}{10000}$).
This was mandatory to have reliable results for the algebraic decay of the correlations.

Four potentials were used for the pair interactions between particules: 
\begin{itemize}
\item Weeks-Chandler-Andersen (WCA) and Lennard-Jones (LJ),
\begin{equation}
U(r)=4u_0 \left[ \left(\frac{\sigma}r\right)^{12} - \left(\frac{\sigma}r\right)^6 \right] \quad r<r_c
\label{eq:WCA}
\end{equation}
To obtain the purely repulsive WCA potential, the value of $r_c$ is fixed at $\sigma 2^{1/6}$. For the classical LJ potential, we consider $r_c\gg \sigma$ to get the attractive part of the potential. It is referred as LJ/cut in LAMMPS.
\item Harmonic,
\begin{equation}
U(r)=\kappa (r_c-r)^2 \quad r<r_c
\label{eq:harmonic}
\end{equation}
In the simulations, we choose $r_c=1$. This potential is referred as harmonic/cut in LAMMPS.

\item Gaussian (gauss/cut in LAMMPS),
\begin{equation}
U(r) = \frac H{\sigma\sqrt{2\pi}} \exp \left[ - \frac{(r-r_m)^2}{2\sigma^2}\right]
\label{eq:gauss}
\end{equation}
The peak position, $r_m$, was fixed at zero.
The global cutoff was chosen to obtain a convergence of the $g(r)$ correlation function computed from the simulation results.
\end{itemize}

In the manuscript, we use $\epsilon$ to describe the magnitude of the interaction potential. For WCA/LJ, harmonic and Gaussian, this parameter is defined as $\epsilon=4u_0, \kappa, H/\sigma\sqrt{2\pi}$, respectively.

\begin{widetext}

\section{Equilibrium calculation and attempts to generalize out-of-equilbrium}

\subsection{Equilibrium}

\subsubsection{Integration of the equilibrium distribution}
\label{sec:eq_distrib}

The equilibrium distribution of three particles interacting via the potential $V(r)=Tv(r)$ is given by
\begin{equation}
  f(\xx_1,\xx_2, \xx_3) \propto \exp\left(-v_{12} - v_{13} - v_{23}\right),
\end{equation}
where we denote $v_{ij}=v(\xx_i-\xx_j)$.
Integrating over $\xx_3$, we obtain the pair correlation up to a constant
\begin{equation}
 g(\xx_1-\xx_2) \propto \int f(\xx_1,\xx_2, \xx_3)\dd\xx_3 \propto \exp\left(-v_{12}\right) \int \exp\left(-v_{13}-v_{23}\right)\dd\xx_3.
\end{equation}
The normalisation factor $Z^{-1}$ should be set such that $g(r)\to 1$ as $r\to\infty$.
When the particles $1$ and $2$ are far appart,
\begin{equation}
  \exp\left(-v_{13}-v_{23}\right)=\exp\left(-v_{13}\right)+\exp\left(-v_{23}\right)-1,
\end{equation}
hence the normalisation factor is
\begin{equation}
  Z=\int\left[1+\exp\left(-v_{13}\right)-1+\exp\left(-v_{23}\right)-1\right]\dd\xx_3.
\end{equation}
Dividing the numerator and denominator by the volume $\mcV$ of the system and using brackets $\langle\cdot\rangle$ for the volume average over $\xx_3$, we find, in the large volume limit
\begin{align}
\ed^{v_{12}} g(\xx_1-\xx_2) &= \frac{1+\mcV^{-1}\int \left(\ed^{-v_{13} -v_{23}}-1\right)\dd\xx_3}{1+\mcV^{-1}\int \left(\ed^{-v_{13}}-1\right)\dd\xx_3+\mcV^{-1}\int \left(\ed^{-v_{23}}-1\right)\dd\xx_3}\\
  &= 1+\mcV^{-1}\int \left(\ed^{-v_{13} -v_{23}}-\ed^{-v_{13}}-\ed^{-v_{23}}+1\right)\dd\xx_3\\
  &= 1+\mcV^{-1}\int \left(\ed^{-v_{13}}-1\right)\left(\ed^{-v_{23}}-1\right)\dd\xx_3\\
  &= 1+\mcV^{-1}\left[\left(\ed^{-v}-1\right)*\left(\ed^{-v}-1\right)\right](\xx_1-\xx_2).
\end{align}
If there are $N$ particles in the system, corresponding to a density $\rho=N/V$, summing the effect of the particles leads to the correction
\begin{equation}\label{eq:correc_eq_distrib}
 \hat g(\xx) = \frac{\ed^{v_{12}(\xx)} g(\xx)-1}{\rho} = \left[\left(\ed^{-v}-1\right)*\left(\ed^{-v}-1\right)\right](\xx).
\end{equation}
This equation is valid up to the first order in the density.

\subsubsection{Potential of mean force}
\label{sec:eq_pmf}

Another way to derive the same result is to consider the partition function of the particle 3 given the position of the particles 1 and 2:
\begin{equation}
  Z_3(\xx_1, \xx_2) = \int \ed^{-v_{13}-v_{23}}\dd \xx_3 = \mcV\left[1+\mcV^{-1}\int \left(\ed^{-v_{13}-v_{23}}-1\right)\dd \xx_3\right].
\end{equation}
The associated free energy is
\begin{equation}
  F_3(\xx_1, \xx_2) = -T\log(Z_3(\xx_1, \xx_2)) \simeq -T\mcV^{-1}\int \left(\ed^{-v_{13}-v_{23}}-1\right)\dd \xx_3,
\end{equation}
where we have taken the large volume limit and discarded the constant term.
Removing the constant part when $|\xx_1-\xx_2|\to\infty$, we get
\begin{equation}
  F_3(\xx_1, \xx_2) = -\frac{T}{\mcV}\left[\left(\ed^{-v}-1\right)*\left(\ed^{-v}-1\right)\right](\xx_1-\xx_2),
\end{equation}
which is the effective potential created by the third particle.
Summing over the particles, we get the potential of mean force to the first order in the density,
\begin{equation}\label{eq:eq_pmf}
  F(\xx_1-\xx_2) = -\rho T\left[\left(\ed^{-v}-1\right)*\left(\ed^{-v}-1\right)\right](\xx_1-\xx_2).
\end{equation}
The pair correlation of the particles $1$ and $2$ is now given by
\begin{equation}\label{eq:correc_eq_pmf}
  \hat g(\xx) = \rho^{-1}\left[\exp\left(-\frac{F(\xx)}{T}\right)-1\right] \simeq \left[\left(\ed^{-v}-1\right)*\left(\ed^{-v}-1\right)\right](\xx),
\end{equation}
where the last equality corresponds to the dilute limit where the correction is small.
The previous result, Eq.~(\ref{eq:correc_eq_distrib}), is recovered.

\subsection{Out of equilibrium tentative generalizations}

In this section, we try to give the ``natural'' generalizations of the two arguments above to the situation where the particles $1$ and $2$ are connected to a thermostat with temperature $T_A$ while the other particles are connected to a different thermostat with temperature $T_B$.

\subsubsection{Integration of the many-body distribution}
\label{sec:noneq_distrib}

The pair correlation of two particles $i$ and $j$ is an effective equilibrium distribution at temperature $T_{ij}=(T_i+T_j)/2$~\cite{Grosberg2015}.
It seems natural to assume that the three-body distribution is given by
\begin{equation}
  f(\xx_1,\xx_2, \xx_3) \propto \exp\left(-\frac{V_{12}}{T_A} - \frac{V_{13} + V_{23}}{T_{AB}}\right).
\end{equation}
Using this distribution in the derivation presented in Sec.~\ref{sec:eq_distrib} leads to
\begin{equation}\label{eq:correc_neq_distrib}
  \hat g(\xx) = \left[\left(\ed^{-V/T_{AB}}-1\right)*\left(\ed^{-V/T_{AB}}-1\right)\right](\xx).
\end{equation}

\subsubsection{Potential of mean force}
\label{sec:noneq_pmf}

Using again the two-body effective equilibrium, the partition function of the particle 3 given the position of the particles 1 and 2 is
\begin{equation}
  Z_3(\xx_1, \xx_2) = \int \ed^{-(V_{13}+V_{23})/T_{AB}}\dd \xx_3.
\end{equation}
Following the derivation of Sec.~\ref{sec:eq_pmf} with this expression, we can consider that the potential of mean force generated by the particles connected to a thermostat with temperature $T_B$ is
\begin{equation}\label{eq:neq_pmf}
  F(\xx_1-\xx_2) = -\rho T_B\left[\left(\ed^{-V/T_{AB}}-1\right)*\left(\ed^{-V/T_{AB}}-1\right)\right](\xx_1-\xx_2).
\end{equation}
Finally, the correction is
\begin{equation}\label{eq:correc_neq_pmf}
  \hat g(\xx) = \frac{T_B}{T_A}\left[\left(\ed^{-V/T_{AB}}-1\right)*\left(\ed^{-V/T_{AB}}-1\right)\right](\xx).
\end{equation}
This result is different from Eq.~(\ref{eq:correc_neq_distrib}).

\section{Linearized Stochastic Density Field Theory}
\label{sec:sdft}

\subsection{From populations of particles to Gaussian fields}

\subsubsection{Dean equation for the density fields}\label{}

The density fields are defined by $\hat\rho_\alpha(\xx,t)=\sum_{x\in I_\alpha}\delta(\xx-\xx_i(t))$, where $I_\alpha$ are the indices of the particles of the species $\alpha$.
The density fields follow the Dean equation:
\begin{equation}\label{eq:dean}
{\dot {\hat{\rho}}}_\alpha=\nabla\cdot \left[T_\alpha\nabla\hat\rho_\alpha+\hat\rho_\alpha \sum_\beta \nabla V*\hat\rho_\beta+\sqrt{2T_\alpha\hat\rho_\alpha}\eeta_\alpha \right],
\end{equation}
where $\eeta_\alpha(\xx,t)$ is a Gaussian white noise:
\begin{equation}
\langle \eeta_\alpha(\xx,t) \eeta_\beta(\xx',t') \rangle=\delta_{\alpha\beta}\delta(\xx-\xx')\delta(t-t').
\end{equation}

\subsubsection{Linearized Dean equation}\label{}

To linearize the Dean equation around the average densities $\rho_\alpha$, we introduce the rescaled density fluctuations $\phi_\alpha(\xx,t)=(\hat\rho_\alpha(\xx,t)-\rho_\alpha)/\sqrt{\rho_\alpha}$ and $v_{\alpha\beta}(\xx)=\sqrt{\rho_\alpha \rho_\beta}V_{\alpha\beta}(\xx)$.
Then we take the limit $\rho_\alpha\to\infty$ while keeping $v_{\alpha\beta}$ constant; the density fluctuations $\phi_\alpha(\xx,t)$ follow
\begin{equation}\label{eq:dean_lin}
\dot\phi_\alpha = \nabla\cdot \left[T_\alpha\nabla\phi_\alpha+ \sum_\beta \nabla v_{\alpha\beta}*\phi_\beta+\sqrt{2T_\alpha}\eeta_\alpha \right],
\end{equation}


\subsubsection{Mapping to a general Gaussian field model}\label{}

We map the model above to a more general model with two Gaussian fields $\phi_\alpha(\xx,t)$ with energy
\begin{equation}
H[\phi_1, \phi_2] = \frac{1}{2}\int \left[\phi_1(\xx)A_1\phi_1(\xx)+\phi_2(\xx)A_2\phi_2(\xx)+ 2\phi_1(\xx)B\phi_2(\xx)\right]\dd\xx,
\end{equation}
where $A_\alpha(\xx)$ and $B(\xx)$ are operators. 
In this section, it is more convenient to use $1$ and $2$ instead of $A$ and $B$ to denote the species.
The overdamped dynamics deriving from this energy is
\begin{align}
\dot\phi_1(\xx,t) & = -R_1[A_1\phi_1(\xx,t)+B\phi_2(\xx,t)]+ \xi_1(\xx,t),\\
\dot\phi_2(\xx,t) & = -R_2[A_2\phi_2(\xx,t)+B\phi_1(\xx,t)] + \xi_2(\xx,t),
\end{align}
where $R_\alpha(\xx)$ are the mobility tensors and the noises $\xi_\alpha(\xx,t)$ have correlations
\begin{equation}
\langle \xi_\alpha(\xx,t)\xi_\beta(\xx',t') \rangle = 2\delta_{\alpha\beta}T_\alpha R_\alpha(\xx-\xx')\delta(t-t'),
\end{equation}
where $T_\alpha$ is the temperature of the field $\alpha$.

The linearized Dean equation (\ref{eq:dean_lin}) corresponds to the following operators in Fourier space,
\begin{align}
\tilde A_\alpha(\kk) & = T_\alpha + \tilde v_{\alpha \alpha}(\kk)\label{eq:mapping_A}\\
\tilde B(\kk) & = \tilde v_{12}(\kk)\\
\tilde R_\alpha(\kk) & = k^2.
\end{align}

The pair correlation of the particles is related to the correlation of the fields $\phi_\alpha$:
\begin{equation}
h_{\alpha\beta}(\xx)
=g_{\alpha\beta}(\xx)-1 
= \left\langle \left(\frac{\hat\rho_\alpha(\xx)}{\rho_\beta}-1 \right)\left(\frac{\hat\rho_\beta(0)}{\rho_j}-1 \right) \right\rangle 
= \frac{C_{\alpha\beta}(\xx)}{\sqrt{\rho_\alpha\rho_\beta}},
\end{equation}
where
\begin{equation}
C_{\alpha\beta}(\xx)=\langle \phi_\alpha(\xx)\phi_\beta(0) \rangle.
\end{equation}

\subsection{Calculation of the correlations}\label{}

\subsubsection{Equations}\label{}

The computations are done in Fourier space, and we define
\begin{equation}
\Phi = \begin{pmatrix}
\phi_1\\\phi_2
\end{pmatrix}.
\end{equation}
It follows
\begin{equation}\label{eq:dotPhi}
\dot\Phi = -\mcR\mcA\Phi + \xi,
\end{equation}
with 
\begin{align}
\mcR & = \begin{pmatrix}
R_1 & 0 \\ 0 & R_2
\end{pmatrix}\\
\mcA & = \begin{pmatrix}
A_1& B \\
B & A_2
\end{pmatrix}.
\end{align}

The correlation $C=\langle \Phi\Phi\transp \rangle$ satisfies in the stationnary state 
\begin{equation}\label{eq:eq_correl}
RAC+CA(-\kk)\transp R(-\kk)\transp = 2TR,
\end{equation}
with $T = \begin{pmatrix}
T_1 & 0 \\ 0 & T_2
\end{pmatrix}$.

\subsubsection{Solution}\label{}

We expand Eq.~(\ref{eq:eq_correl}) and get
\begin{align}
A_1C_1 + \frac{B}{2}(C_{12}+C_{21}) & = T_1, \label{eq:11}\\
\Gamma C_{12} + R_2 B C_1 + R_1 B C_2 & = 0, \label{eq:12}\\
\Gamma C_{21} + R_2 B C_1 + R_1 B C_2 & = 0, \label{eq:21}\\
A_2C_2 + \frac{B}{2}(C_{12}+C_{21}) & = T_2. \label{eq:22}
\end{align}
where we have defined
\begin{equation}
\Gamma = R_1A_1+R_2A_2.
\end{equation}

From Eqs.~(\ref{eq:11}, \ref{eq:22}), we get
\begin{equation}
A_2C_2- A_1C_1 = T_2- T_1,
\end{equation}
hence 
\begin{equation}\label{eq:c2_c1}
C_2 = \frac{T_2- T_1}{A_2} +  \frac{A_1}{A_2}C_1.
\end{equation}
From Eqs.~(\ref{eq:12}, \ref{eq:21}), we have
\begin{equation}
\frac{C_{12}+C_{21}}{2} = -\frac{ B}{\Gamma}( R_2C_1+R_1C_2)
=-\frac{ B}{\Gamma A_2}\left[\Gamma C_1 + R_1(T_2- T_1) \right],
\end{equation}
where we have used Eq.~(\ref{eq:c2_c1}).
Using this relation in Eq.~(\ref{eq:11}), we get
\begin{equation}
A_1 C_1 - \frac{ B^2}{\Gamma A_2}\left[\Gamma C_1 + R_1(T_2- T_1) \right] = T_1,
\end{equation}
which leads to
\begin{equation}\label{eq:c1}
C_1 = \frac{T_1 \Gamma  A_2  + (T_2- T_1) R_1 B^2}{\Gamma [A_1A_2-B^2]}.
\end{equation}
Using this relation in Eq.~(\ref{eq:c2_c1}), we get
\begin{equation}\label{eq:c2}
C_2 = \frac{T_2\Gamma A_1 - (T_2- T_1) R_2 B^2}{\Gamma[A_1A_2-B^2]}.
\end{equation}
Finally, using these relations in Eqs.~(\ref{eq:12}, \ref{eq:21}), we obtain
\begin{equation}\label{eq:c12}
C_{12} = - \frac{B(T_2 R_1 A_1 +  T_1R_2A_2)}{\Gamma[A_1A_2-B^2]}
\end{equation}
and $C_{21}=C_{12}^*$.

\subsubsection{Equilibrium}\label{}

If the system is at equilibrium, $T_1=T_2=T$, the correlations are
\begin{equation}
C = \frac{T}{A_1A_2-B^2}\begin{pmatrix}
A_2 & -B \\ -B & A_1
\end{pmatrix}
= T\mcA^{-1},
\end{equation}
which is the expected result.

\subsection{Application}\label{}

\subsubsection{Pair potential in Fourier space}

In our case, we use harmonic spheres, corresponding to $V(r)=\epsilon(1-r)^2\theta(1-r)$, $\theta$ being the Heaviside function.
The Fourier transform of the potential reads
\begin{equation}
\tilde V(\kk) = \int e^{-i\kk\cdot\xx} V(\xx)\dd\xx = \epsilon\pi\int_0^1 x(1-x)^2J_0(kx)\dd x
= \frac{\epsilon\pi}{k^2} \left[\pi H_0(k)J_1(k)-\pi H_1(k)J_0(k)-2J_2(k) \right],
\end{equation}
where $J_n(z)$ is the Bessel function of the first kind and $H_n(z)$ is the Struve function.

The pair correlation function $h_{11}(r)$ is plotted in Fig.~\ref{fig:sdft} for different values of the temperature $T_2$: a good agreement is obtained, both at equilibrium and out of equilibrium.

\begin{figure}
\begin{center}
\includegraphics[width=.5\linewidth]{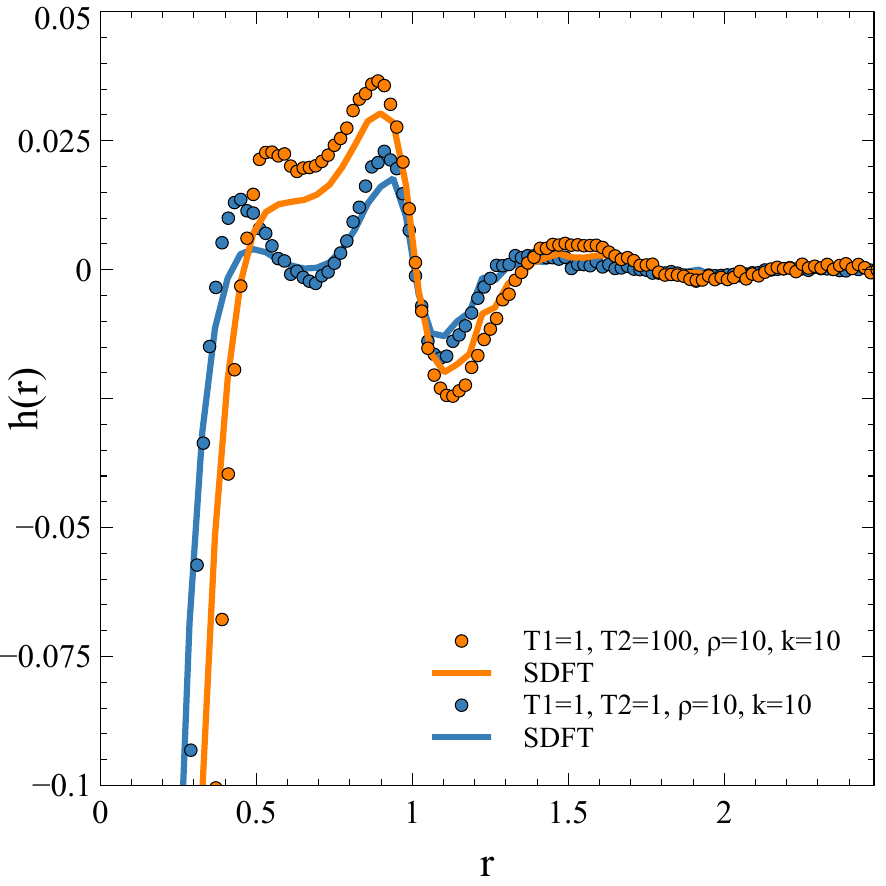}
\end{center}
\caption{Pair correlation function $h(r)=g(r)-1$ of particles of the species 1, with $\rho=10$, $T_1=1$, for different temperatures $T_2$.}
\label{fig:sdft}
\end{figure}

\subsubsection{Low density}\label{}

In the low density limit, we find
\begin{equation}
\label{eq:h1_sdft_dilute}
  \tilde h_{11} = \frac{\tilde C_1-1}{\rho_1} \simeq -\frac{\tilde V}{T_1}+\sum_\alpha  \frac{\rho_\alpha}{T_1 T_{1\alpha}} \tilde V^2,
\end{equation}
where $T_{\alpha\beta} = (T_\alpha+T_\beta)/2$.

Going back to real space, denoting the species $A$ and $B$ and setting $\rho_A=0$, $\rho_B=\rho$, this relation reads
\begin{equation}
  h(r) = -\frac{V(r)}{T_A}+\frac{\rho}{T_A T_{AB}} V*V(r).
\end{equation}
The correlation $\hat g(r)$ that we focus on is
\begin{align}
  \hat g(r) &= \frac{e^{V(r)/T_A}\left[1+h(r)\right]-1}{\rho}\\
  &=\frac{1}{\rho}\left(e^{V(r)/T_A}\left[1-\frac{V(r)}{T_A}+\frac{\rho}{T_A T_{AB}} V*V(r)\right]-1\right)\\
  &=\frac{V*V(r)}{T_A T_{AB}}.
\end{align}
To obtain the last line, we have eliminated the terms of order $\epsilon^2$ that do not involve the density $\rho$. This is justified by the fact that the linearization of SDFT amounts to take $\epsilon\to 0$ while keeping $\rho\epsilon$ constant.

\section{Low density expansion}\label{}

\subsection{Smoluchowski equation and many-body potential}
\label{sec:mbp}

The $N$-particle distribution $f(\XX,t)$, where $\XX=(\xx_1,\dots\xx_N)$, evolves according to the Smoluchowski equation $\partial_t f=\mcL f$, where $\mcL$ is the Liouville operator:
\begin{equation}\label{eq:liouv0}
\mcL f(\XX) = \sum_i\nabla_i\cdot \left[T_i\nabla_i f(\XX) + f(\XX)\sum_{j\neq i}\nabla V_{ij}(\xx_i-\xx_j) \right].
\end{equation}
We are looking for the stationnary distribution, which satisfies $\mcL f=0$.

We write the distribution under the form
\begin{equation}
f(\XX) = \exp(-\psi(\XX)).
\end{equation}
Using that $\nabla_i f = -f\nabla_i\psi$ and $\nabla_i^2 f=f[(\nabla_i\psi)^2-\nabla_i^2\psi]$, we find
\begin{equation}\label{eq:liouv_gen}
\frac{\mcL f(\XX)}{f(\XX)} = \sum_i T_i \left[(\nabla_i\psi)^2-\nabla_i^2\psi \right]+\sum_{\langle ij \rangle} T_{ij}\left[2\nabla^2 v_{ij}-\nabla v_{ij}\cdot(\nabla_i\psi-\nabla_j\psi) \right],
\end{equation}
where $v_{ij}=V_{ij}/T_{ij}$.

We separate the two-particle contribution, which defines the many-body potential $\phi(\XX)$:
\begin{equation}
\psi(\XX)=\sum_{\langle ij \rangle} v_{ij} + \phi(\XX),
\end{equation}
Using that
\begin{align}
\nabla_i\psi & = \sum_{j\neq i}\nabla v_{ij} + \nabla_i\phi,\\
(\nabla_i\psi)^2 & = \sum_{j,k\neq i}\nabla v_{ij}\cdot\nabla v_{ik} + 2 \sum_{j\neq i}\nabla v_{ij} \cdot \nabla_i\phi + (\nabla_i\phi)^2,\\
\nabla_i^2\psi & = \sum_{j\neq i}\nabla^2 v_{ij} + \nabla_i^2\phi,
\end{align}
the general expression (\ref{eq:liouv_gen}) becomes
\begin{multline}
\frac{\mcL f(\XX)}{f(\XX)} = \sum_i T_i \left[\sum_{j,k\neq i}\nabla v_{ij}\cdot\nabla v_{ik} + 2 \sum_{j\neq i}\nabla v_{ij} \cdot \nabla_i\phi + (\nabla_i\phi)^2-\sum_{j\neq i}\nabla^2 v_{ij} - \nabla_i^2\phi \right]\\+\sum_{\langle ij \rangle} T_{ij}\left[2\nabla^2 v_{ij}-\nabla v_{ij}\cdot\left(\sum_{k\neq i}\nabla v_{ik}-\sum_{l\neq j}\nabla v_{jl}+\nabla_i\phi-\nabla_j\phi\right) \right],
\end{multline}
The terms in $\nabla^2v_{ij}$ cancel.
Moreover, the terms in $\nabla v_{ij}\cdot\nabla v_{ik}$ and $\nabla v_{ij}\cdot\nabla v_{jl}$ cancel if $k=j$ and $l=i$: only the terms involving three different particles remain:
\begin{multline}
\frac{\mcL f(\XX)}{f(\XX)} = \sum_i T_i \left[2\sum_{\langle jk \rangle\notowner i}\nabla v_{ij}\cdot\nabla v_{ik} + 2 \sum_{j\neq i}\nabla v_{ij} \cdot \nabla_i\phi + (\nabla_i\phi)^2- \nabla_i^2\phi \right]\\+\sum_{\langle ij \rangle} T_{ij}\left[-\nabla v_{ij}\cdot\left(\sum_{k\notin \langle ij \rangle}\nabla v_{ik}-\sum_{k\notin \langle ij \rangle}\nabla v_{jk}+\nabla_i\phi-\nabla_j\phi\right) \right].
\end{multline}

This can be written as a sum over the particles, a sum over the pairs and a sum over the triplets:
\begin{multline}
\frac{\mcL f(\XX)}{f(\XX)} = \sum_i T_i \left[(\nabla_i\phi)^2- \nabla_i^2\phi \right] + \sum_{\langle ij \rangle} \nabla v_{ij}\cdot \left[(2T_i-T_{ij})\nabla_i\phi - (2T_j-T_{ij})\nabla_j\phi \right]\\
 +\sum_{\langle ijk \rangle} \left[2T_i \nabla v_{ij}\cdot\nabla v_{ik} + 2T_j \nabla v_{ji}\cdot\nabla v_{jk}+2T_k\nabla v_{ki}\cdot\nabla v_{kj}-T_{ij}\nabla v_{ij}\cdot(\nabla v_{ik}-\nabla v_{jk})\right.\\
 \left.-T_{jk}\nabla v_{jk}\cdot(\nabla v_{ji}-\nabla v_{ki})-T_{ki}\nabla v_{ki}\cdot(\nabla v_{kj}-\nabla v_{ij}) \right].
\end{multline}
The triplets terms can be rearranged:
\begin{multline}
\frac{\mcL f(\XX)}{f(\XX)} = \sum_i T_i \left[(\nabla_i\phi)^2- \nabla_i^2\phi \right] + \sum_{\langle ij \rangle} \nabla v_{ij}\cdot \left[(2T_i-T_{ij})\nabla_i\phi - (2T_j-T_{ij})\nabla_j\phi \right]\\
 +\sum_{\langle ijk \rangle} \left[(2T_i-T_{ij}-T_{ik}) \nabla v_{ij}\cdot\nabla v_{ik} + (2T_j-T_{ij}-T_{jk}) \nabla v_{ji}\cdot\nabla v_{jk}+(2T_k-T_{ik}-T_{jk})\nabla v_{ki}\cdot\nabla v_{kj} \right].
\end{multline}
We can then simplify the combination of temperatures:
\begin{align}
2T_i-T_{ij} & =\frac{3T_i-T_j}{2}=T_{ij}+\tau_{ij},\\
2T_i-T_{ij}-T_{ik}& = T_i-T_{jk}=\tau_{ijk},
\end{align}
where we have introduced the temperature differences
\begin{align}
    \tau_{ij} &= T_i-T_j,\\
    \tau_{ijk} &=T_i-T_{jk}.
    \label{eq:def_tau}
\end{align}
Finally, we have
\begin{multline}\label{eq:liouv_gen3}
\frac{\mcL f(\XX)}{f(\XX)} = \sum_i T_i \left[(\nabla_i\phi)^2- \nabla_i^2\phi \right] + \sum_{\langle ij \rangle} \nabla v_{ij}\cdot \left[(T_{ij}+\tau_{ij})\nabla_i\phi - (T_{ij}-\tau_{ij})\nabla_j\phi \right]\\
+\sum_{\langle ijk \rangle} \left[\tau_{ijk} \nabla v_{ij}\cdot\nabla v_{ik} +\tau_{jki} \nabla v_{ji}\cdot\nabla v_{jk}+\tau_{kij}\nabla v_{ki}\cdot\nabla v_{kj} \right].
\end{multline}

\subsection{Perturbative expansion}\label{}

\subsubsection{Order $\epsilon^2$}
\label{sec:mbp2}

We solve Eq.~(\ref{eq:liouv_gen3}) perturbatively in the pair interaction $v\propto\epsilon$: we write $\phi=\sum_{k=2}^\infty \phi^{(k)}$, where $\phi^{(k)}\propto\epsilon^k$.
The lowest order is $\phi^{(2)}$, which is the solution of
\begin{equation}\label{eq:liouv_phi3_weak}
    \sum_i T_i \nabla_i^2\phi^{(2)} =  \sum_{\langle ijk \rangle} \left[\tau_{ijk} \nabla v_{ij}\cdot\nabla v_{ik} +\tau_{jki} \nabla v_{ji}\cdot\nabla v_{jk}+\tau_{kij}\nabla v_{ki}\cdot\nabla v_{kj} \right].
\end{equation}
This is an anisotropic Poisson equation. We can write the solution as
\begin{equation}\label{eq:phi_triplets}
    \phi^{(2)} = \sum_{\langle ijk \rangle}\tau_{ijk}w^{(2)}_{ijk}+\tau_{jki}w^{(2)}_{jki}+\tau_{kij}w^{(2)}_{kij},
\end{equation}
where $w^{(2)}_{ijk}$ is the solution of
\begin{equation}
\sum_l T_l \nabla_l^2 w^{(2)}_{ijk} =   \nabla v_{ij}\cdot\nabla v_{ik}.
\end{equation}

First, we expect $w^{(2)}_{ijk}$ to depend only on the coordinates of the particles $i$, $j$ and $k$.
Moreover, we expect it to depend only on the particle separations $\yy=\xx_j-\xx_i$ and $\yy'=\xx_k-\xx_i$:
\begin{equation}\label{eq:wijk}
w^{(2)}_{ijk}(\xx_i,\xx_j,\xx_k) = \omega^{(2)}_{ijk}(\xx_j-\xx_i,\xx_k-\xx_i),
\end{equation}
which is solution of
\begin{equation}
2\left(T_{ij}\nabla^2+T_{ik}\nabla'^2+T_i\nabla\cdot\nabla'\right) \omega^{(2)}_{ijk}(\yy,\yy') =   \nabla v_{ij}(\yy)\cdot\nabla v_{ik}(\yy').
\end{equation}
In Fourier space, it becomes
\begin{equation}
2\left(T_{ij}k^2+T_{ik}k'^2+T_i\kk\cdot\kk'\right) \tilde\omega^{(2)}_{ijk}(\kk,\kk') =   \kk\cdot\kk' \tilde v_{ij}(\kk)\tilde v_{ik}(\kk'),
\end{equation}
which is solved by
\begin{equation}\label{eq:om_ijk_fourier}
\tilde\omega^{(2)}_{ijk}(\kk,\kk') = \frac{\kk\cdot\kk' \tilde v_{ij}(\kk)\tilde v_{ik}(\kk')}{2\left(T_{ij}k^2+T_{ik}k'^2+T_i\kk\cdot\kk'\right)}.
\end{equation}

\subsubsection{Order $\epsilon^3$}
\label{sec:mbp3}

The next order is $\phi^{(3)}$, which is the solution of
\begin{equation}\label{eq:phi33}
\sum_i T_i \nabla_i^2\phi^{(3)}  =  \sum_{\langle ij \rangle} \nabla v_{ij}\cdot \left[(T_{ij}+\tau_{ij})\nabla_i\phi^{(2)} - (T_{ij}-\tau_{ij})\nabla_j\phi^{(2)} \right].
\end{equation}
Using the expression of $\phi^{(2)}$ (Eq.~\eqref{eq:phi_triplets}), it becomes
\begin{equation}\label{eq:phi33_triplets}
\sum_i T_i \nabla_i^2\phi^{(3)}  =  \sum_{\langle ij \rangle} \nabla v_{ij}\cdot \left[T_{ij}(\nabla_i-\nabla_j)+\tau_{ij}(\nabla_i + \nabla_j)\right]\sum_{\langle klm \rangle} (\tau_{klm}w^{(2)}_{klm}+\tau_{lmk}w^{(2)}_{lmk}+\tau_{mkl}w^{(2)}_{mkl}).
\end{equation}

The double sum in Eq.~\eqref{eq:phi33_triplets} can be simplified: it is 0 if neither $i$ nor $j$ belongs to $\langle klm \rangle$.
If only $i$ or $j$ belongs to $\langle klm\rangle$, the correction to the pair correlation will be of order $\rho^2$ and will not contribute to $g_1(r)$. 
At order $\rho$, we only have to consider three particles, so that there is only one triplet and three pairs.
We write again
\begin{equation}
\phi^{(3)} = \tau_{123}w_{123}^{(3)}+\tau_{231}w_{231}^{(3)}+\tau_{312}w_{312}^{(3)},
\end{equation}
and the first term, for instance, is solution of
\begin{multline}
  \sum_i T_i\nabla_i^2 w_{123}^{(3)} =  \left(
  \nabla v_{12}\cdot\left[T_{12}(\nabla_1-\nabla_2)+\tau_{12}(\nabla_1+\nabla_2)\right]
  +\nabla v_{13}\cdot\left[T_{13}(\nabla_1-\nabla_3)+\tau_{13}(\nabla_1+\nabla_3)\right]\right.\\
  \left.+\nabla v_{23}\cdot\left[T_{23}(\nabla_2-\nabla_3)+\tau_{23}(\nabla_2+\nabla_3)\right]
  \right)w^{(2)}_{123}.
\end{multline}

Writing again
\begin{equation}
w_{123}^{(3)} = w_{123}^{(3)}(\xx_1,\xx_2,\xx_3) = \omega_{123}^{(3)} (\xx_2-\xx_1,\xx_3-\xx_1),
\end{equation}
we find that the equation for $\omega_{123}^{(3)}(\yy,\yy')$ reads
\begin{multline}
2\left(T_{12}\nabla^2+T_{13}\nabla'^2+T_1\nabla\cdot\nabla'\right)\omega_{123}^{(3)}(\yy,\yy') 
\\= \left(\nabla v_{12}(\yy)\cdot\left[T_{12}(2\nabla+\nabla')+\tau_{12}\nabla'\right]+\nabla v_{13}(\yy')\cdot\left[T_{13}(\nabla+2\nabla')+\tau_{13}\nabla\right]\right.
\\\left.+\nabla v_{23}(\yy-\yy')\cdot\left[T_{23}(\nabla-\nabla')+\tau_{23}(\nabla+\nabla')\right]\right)\omega_{123}(\yy,\yy').
\end{multline}
Fourier transforming, we arrive at
\begin{multline}
\tilde \omega_{123}^{(3)}(\kk,\kk') = \frac{1}{2(T_{12}k^2+T_{13}k'^2+T_1\kk\cdot\kk')}
\int \frac{\dd\kk''}{(2\pi)^d} \\
\left((\kk-\kk'')\cdot\left[T_{12}(2\kk''+\kk')+\tau_{12}\kk'\right]\tilde v_{12}(\kk-\kk'')\tilde\omega_{123}^{(2)}(\kk'',\kk')\right.\\
+(\kk'-\kk'')\cdot\left[T_{13}(\kk+2\kk'')+\tau_{13}\kk\right]\tilde v_{13}(\kk'-\kk'')\tilde\omega_{123}^{(2)}(\kk,\kk'')\\
\left.+\kk''\cdot\left[T_{23}(\kk-\kk'-2\kk'')+\tau_{23}(\kk+\kk')\right]\tilde v_{23}(\kk'')\tilde\omega_{123}^{(2)}(\kk-\kk'',\kk'+\kk'')\right).
\end{multline}

\subsection{Correction to the 2-body pair correlation}

\subsubsection{General expression}
\label{sec:correc_correl_gen}

The pair correlation between particles 1 and 2 is given by
\begin{equation}
    g_{12}\propto \int f(\XX)\prod_{i>2}\dd\xx_i
    =\int\exp\left(-\sum_{\langle ij\rangle}v_{ij}-\phi(\XX)\right)\prod_{i>2}\dd\xx_i.
    \label{}
\end{equation}
We can take out the direct interaction $v_{12}$ to get the cavity distribution function~\cite{Hansen2006}.
Moreover, assuming that all the particles for $i>2$ are identical and with a density $\rho$, at first order in the density it suffices to consider a 3 particles system and
\begin{equation}
    e^{v_{12}}g_{12}\propto \int\exp\left(-(v_{13}+v_{23})-\phi(\XX)\right)\dd\xx_3.
    \label{}
\end{equation}
There should be a normalization factor such that $g_{12}\to 1$ as $|\xx_1-\xx_2|\to\infty$.
In this limit,
\begin{equation}
Z=\int e^{-v_{13}-v_{23}-\phi}\dd\xx_3
=\int \left[1+(e^{-v_{12}}-1)+(e^{-v_{23}}-1)\right]\dd\xx_3,
\end{equation}
hence
\begin{align}
    e^{v_{12}}g_{12}&=\frac{\int e^{-v_{13}-v_{23}-\phi}\dd\xx_3}{\int \left[1+(e^{-v_{12}}-1)+(e^{-v_{23}}-1)\right]\dd\xx_3}\\
    &=\frac{\left\langle e^{-v_{13}-v_{23}-\phi}\right\rangle}{1+\left\langle e^{-v_{13}}-1\right\rangle+\left\langle e^{-v_{23}}-1\right\rangle}\\
    &=\left\langle e^{-v_{13}-v_{23}-\phi}\right\rangle-\left\langle e^{-v_{13}}-1\right\rangle-\left\langle e^{-v_{23}}-1\right\rangle.
    \label{}
\end{align}
We see that the two negative terms cancel the terms where only one of the two particles appears in the first term.
Finally, summing over the particles with a density $\rho$ leads to
\begin{equation}
    \hat g(\xx_1-\xx_2) = \frac{e^{v_{12}}g_{12}-1}{\rho}=\int\left[ e^{-v_{13}-v_{23}-\phi}- e^{-v_{13}}-e^{-v_{23}}+1 \right]\dd\xx_3.
    \label{eq:correc_g12}
\end{equation}

Now we can expand the term in the integrals in powers of the potential, up to $v^3$:
\begin{equation}
    \hat g(\xx_1-\xx_2)=\int \left[v_{13}v_{23}-\phi^{(2)}-\frac{1}{2}\left(v_{13}^2v_{23}+v_{13}v_{23}^2\right)+(v_{13}+v_{23})\phi^{(2)}-\phi^{(3)}\right]\dd\xx_3.
    \label{eq:correc_g12_expansion}
\end{equation}
The first two terms in the integrand scale as $\epsilon^2$, the last three terms scale as $\epsilon^3$.

\subsubsection{Order $\epsilon^2$}
\label{sec:correc_correl_2}

\paragraph{First term.}

The first term in the integrand of Eq.~\eqref{eq:correc_g12_expansion} leads to the correction
\begin{equation}
    \hat g_{2,1}=\bar\rho v_{13}*v_{23}.
    \label{eq:correc_from_2body}
\end{equation}

\paragraph{Second term.}\label{}

The second term in the integrand of Eq.~\eqref{eq:correc_g12_expansion} contributes
\begin{equation}
    \hat g_{2,2}=-\int\phi^{(2)}\dd\xx_3
    \label{}
\end{equation}
Using the decomposition (\ref{eq:phi_triplets}), we have to determine the integral of $w^{(2)}_{123}$, $w^{(2)}_{231}$ and $w^{(2)}_{312}$ over $\xx_3$:
\begin{align}
\int w^{(2)}_{123}\dd\xx_3 & = \int\omega^{(2)}_{123}(\xx_2-\xx_1,\xx_3-\xx_1)\dd\xx_3\\
& = \int \frac{\dd\kk\dd\kk'}{(2\pi)^{2d}}\tilde \omega^{(2)}_{123}(\kk,\kk')\int e^{i\kk\cdot(\xx_2-\xx_1)+i\kk'\cdot(\xx_3-\xx_1)}\dd\xx_3\\
& = \int \frac{\dd\kk}{(2\pi)^{d}}\tilde \omega^{(2)}_{123}(\kk,0) e^{i\kk\cdot(\xx_2-\xx_1)} \label{eq:int_w123_dx3}\\
& = 0.
\end{align}
The second term is
\begin{align}
\int w^{(2)}_{231}\dd\xx_3 & = \int\omega^{(2)}_{231}(\xx_3-\xx_2,\xx_1-\xx_2)\dd\xx_3\\
& = \int \frac{\dd\kk\dd\kk'}{(2\pi)^{2d}}\tilde \omega^{(2)}_{231}(\kk,\kk')\int e^{i\kk\cdot(\xx_3-\xx_2)+i\kk'\cdot(\xx_1-\xx_2)}\dd\xx_3\\
& = \int \frac{\dd\kk'}{(2\pi)^{d}}\tilde \omega^{(2)}_{231}(0,\kk') e^{i\kk'\cdot(\xx_1-\xx_2)} \label{eq:int_w231_dx3}\\
& = 0.
\end{align}
The third term is
\begin{align}
\int w^{(2)}_{312}\dd\xx_3 & = \int\omega^{(2)}_{312}(\xx_1-\xx_3,\xx_2-\xx_3)\dd\xx_3\\
& = \int \frac{\dd\kk\dd\kk'}{(2\pi)^{2d}}\tilde \omega^{(2)}_{312}(\kk,\kk')\int e^{i\kk\cdot(\xx_1-\xx_3)+i\kk'\cdot(\xx_2-\xx_3)}\dd\xx_3\\
& = \int \frac{\dd\kk}{(2\pi)^{d}}\tilde \omega^{(2)}_{312}(\kk,-\kk) e^{i\kk\cdot(\xx_1-\xx_2)} \label{eq:int_w312_dx3}\\
& = -\frac{1}{2T_{12}}\int \frac{\dd\kk}{(2\pi)^{d}} \tilde v_{13}(\kk)\tilde v_{23}(\kk) e^{i\kk\cdot(\xx_1-\xx_2)}\\
& = -\frac{1}{2T_{12}} v_{13}*v_{23}(\xx_1-\xx_2).
\end{align}
The correction is thus
\begin{equation}
\hat g_{2,2}=-\int\phi^{(2)}\dd\xx_3 = -\tau_{312}\int w^{(2)}_{312}\dd\xx_3 = \frac{\tau_{312}}{2T_{12}}v_{31}*v_{32}(\xx_1-\xx_2).
\end{equation}

If the particles $1$ and $2$ are of type $A$, and the others are of type $B$, summing with the contribution (\ref{eq:correc_from_2body}), we get
\begin{equation}\label{eq:correc_2_combined}
\hat g_2=\hat g_{2,1}+\hat g_{2,2}= \left(\frac{1}{T_{AB}^2}+\frac{T_B-T_A}{2 T_A T_{AB}^2} \right)V*V
= \frac{V*V}{T_A T_{AB}}.
\end{equation}

\subsubsection{Order $\epsilon^3$}
\label{sec:correc_correl_3}

At order $\epsilon^3$, there are three contributions in the expansion (\ref{eq:correc_g12_expansion}).

\paragraph{First term.}

The first term contributes
\begin{equation}\label{eq:y31}
\hat g_{3,1}(\xx)=-\frac{1}{2} \left(v_{13}^2*v_{23} + v_{13}*v_{23}^2\right)(\xx).
\end{equation}

\paragraph{Second term.}
\label{par:y32}

The second term is
\begin{equation}
\hat g_{3,2} = \int (v_{13}+v_{23})\phi^{(2)}\dd\xx_3.
\end{equation}
Using the decomposition (\ref{eq:phi_triplets}), we have to determine the integral of the six terms $v_{13}w_{123}^{(2)}$, etc. over $\xx_3$.
They are (we use $\xx=\xx_1-\xx_2$)
\begin{align}
\int v_{13}w^{(2)}_{123}\dd\xx_3 & 
= \int v_{13}(\xx_1-\xx_3)\omega^{(2)}_{123}(\xx_2-\xx_1,\xx_3-\xx_1)\dd\xx_3\\
& = \int \frac{\dd\kk\dd\kk'}{(2\pi)^{2d}}  \tilde v_{13}(\kk')\tilde \omega^{(2)}_{123}(\kk,\kk') e^{i\kk\cdot\xx}.
\end{align}
The second is
\begin{align}
\int v_{23}w^{(2)}_{123}\dd\xx_3 & 
= \int v_{23}(\xx_2-\xx_3)\omega^{(2)}_{123}(\xx_2-\xx_1,\xx_3-\xx_1)\dd\xx_3\\
& = \int \frac{\dd\kk\dd\kk'}{(2\pi)^{2d}} \tilde v_{23}(\kk)\tilde\omega^{(2)}_{123}(\kk',\kk) e^{i(\kk+\kk')\cdot \xx}\\
& = \int \frac{\dd\kk\dd\kk'}{(2\pi)^{2d}} \tilde v_{23}(\kk')\tilde\omega^{(2)}_{123}(\kk-\kk',\kk') e^{i\kk\cdot \xx}.
\end{align}
The third is
\begin{align}
\int v_{13}w^{(2)}_{231}\dd\xx_3 & 
= \int v_{13}(\xx_1-\xx_3)\omega^{(2)}_{231}(\xx_3-\xx_2,\xx_1-\xx_2)\dd\xx_3\\
& = \int \frac{\dd\kk\dd\kk'}{(2\pi)^{2d}} \tilde v_{13}(\kk)\tilde\omega^{(2)}_{231}(\kk,\kk') e^{i(\kk+\kk')\cdot \xx}\\
& = \int \frac{\dd\kk\dd\kk'}{(2\pi)^{2d}}\tilde v_{13}(\kk')\tilde\omega^{(2)}_{231}(\kk',\kk-\kk') e^{i\kk\cdot \xx}.
\end{align}
The fourth is
\begin{align}
\int v_{23}w^{(2)}_{231}\dd\xx_3 & 
= \int v_{23}(\xx_2-\xx_3)\omega^{(2)}_{231}(\xx_3-\xx_2,\xx_1-\xx_2)\dd\xx_3\\
& = \int \frac{\dd\kk\dd\kk'}{(2\pi)^{2d}} \tilde v_{23}(\kk')\tilde\omega^{(2)}_{231}(\kk',\kk) e^{i\kk\cdot \xx}.
\end{align}
The fifth is
\begin{align}
\int v_{13}w^{(2)}_{312}\dd\xx_3 & 
= \int v_{13}(\xx_1-\xx_3)\omega^{(2)}_{312}(\xx_1-\xx_3,\xx_2-\xx_3)\dd\xx_3\\
& = \int \frac{\dd\kk\dd\kk'}{(2\pi)^{2d}} \tilde v_{13}(\kk)\tilde\omega^{(2)}_{312}(\kk',-\kk-\kk') e^{i(\kk+\kk')\cdot \xx}\\
& = \int \frac{\dd\kk\dd\kk'}{(2\pi)^{2d}}\tilde v_{13}(\kk')\tilde\omega^{(2)}_{312}(-\kk-\kk',\kk) e^{i\kk\cdot \xx}.
\end{align}
Finally, the sixth is
\begin{align}
\int v_{23}w^{(2)}_{312}\dd\xx_3 & 
= \int v_{23}(\xx_2-\xx_3)\omega^{(2)}_{312}(\xx_1-\xx_3,\xx_2-\xx_3)\dd\xx_3\\
& = \int \frac{\dd\kk\dd\kk'}{(2\pi)^{2d}} \tilde v_{23}(\kk)\tilde\omega^{(2)}_{312}(\kk',-\kk-\kk') e^{i\kk'\cdot \xx}\\
& = \int \frac{\dd\kk\dd\kk'}{(2\pi)^{2d}}\tilde v_{23}(\kk')\tilde\omega^{(2)}_{312}(\kk,-\kk-\kk') e^{i\kk\cdot \xx}.
\end{align}

\paragraph{Third term.}
\label{par:y33}

The third $\epsilon^3$ term in the integrand of Eq.~(\ref{eq:correc_g12_expansion}) is
\begin{equation}
\hat g_{3,3}(r) = -\int \phi^{(3)}\dd\xx_3
=-\int\left(\tau_{123}w_{123}^{(3)}+\tau_{231}w_{231}^{(3)}+\tau_{312}w_{312}^{(3)}\right)\dd\xx_3.
\end{equation}

These integrals have been computed above, they read (Eqs.~(\ref{eq:int_w123_dx3}), (\ref{eq:int_w231_dx3}), (\ref{eq:int_w312_dx3})):
\begin{align}
\int w_{123}^{(3)}\dd\xx_3 & = \int \frac{\dd\kk}{(2\pi)^{d}}\tilde \omega_{123}^{(3)}(\kk,0) e^{i\kk\cdot\xx},\\
\int w_{231}^{(3)}\dd\xx_3 & = \int \frac{\dd\kk}{(2\pi)^{d}}\tilde \omega_{231}^{(3)}(0,\kk) e^{i\kk\cdot\xx},\\
\int w_{312}^{(3)}\dd\xx_3 & = \int \frac{\dd\kk}{(2\pi)^{d}}\tilde \omega_{312}^{(3)}(\kk,-\kk) e^{i\kk\cdot\xx}.
\end{align}

\end{widetext}

\subsection{Scaling relations and numerical integration}

\subsubsection{Scaling form}

All the terms that involve the many-body potential $\phi$ at order $\epsilon^3$ consist of a double integral in Fourier space, three interaction potentials and a scale-invariant combination of the wavevectors (which scales as $k^0$).
While the prefactor is difficult to obtain analytically, this form suggests an algebraic decay as $r^{-2d}$, where $d$ is the spatial dimension, and a prefactor scaling as $\tilde V(0)^3$, where $\tilde V(0)=\int V(\xx)\dd\xx$.
This is confirmed by the numerical integration for the Harmonic and Gaussian potentials presented in Sec.~\ref{sec:num_int} (Fig.~\ref{fig:comp_harm_gauss}).

The last point can be made rigorous if the potential $V(r)$ is rescaled by a factor $\lambda$, which we denote $V_\lambda(r)=\lambda^d V(\lambda r)$, such that $\tilde V_\lambda(k)=\tilde V(\lambda k)$.
It can be proven from the relations above that if the potential $V(r)$ leads to the correction $f(r)$, then the potential $V_\lambda(r)$ leads to the correction $f_\lambda(r)=\lambda^{-2d}f(r/\lambda)$.
As a consequence, if $f(r)\sim a r^{-2d}$, then $f_\lambda(r)\sim \lambda^{-2d}a (r/\lambda)^{-2d}=ar^{-2d}$: the prefactor of the algebraic decay is the same for the initial potential and the rescaled potential.

The dependence in the potential should be accompagned with a dependence in the temperatures, as only the ratio matters, as can be seen from the definition of the Liouville operator (Eq.~\eqref{eq:liouv0}).
If the particles 1 and 2 are of type $A$ and the others are of type $B$, these arguments leads to an asymptotic form
\begin{equation}\label{eq:gen_pert}
  \hat g(r) \underset{r\to\infty}{\sim} \left[\frac{\tilde V(0)}{T_A}\right]^3 F\left(\frac{T_B}{T_A}\right)r^{-2d}.
\end{equation}
As the correction originates from temperature differences, the temperature dependence should satisfy
\begin{equation}
  F(\theta)\underset{\theta\to 1}{\sim}\theta-1.
\end{equation}
Moreover counting the occurence of the temperatures $T_A$ and $T_B$ in the different terms entering the correction, we find that for large temperature differences,
\begin{equation}
  F(\theta)\underset{\theta\to \infty}{\sim}\theta^{-2}.
\end{equation}

\subsubsection{Numerical integration}
\label{sec:num_int}

The induced interaction at order $\epsilon^2$, Eq.~\eqref{eq:correc_2_combined}, involves an integral in real space, which can be evaluated numerically.
The same applies to the first term at order $\epsilon^3$, Eq.~\eqref{eq:y31}.
The two other terms at order $\epsilon^3$ (\ref{par:y32} and \ref{par:y33}) require a double numerical integration in Fourier space and a numerical integration in real space to invert the Fourier transform, which take much more time.

\begin{figure}
\begin{center}
\includegraphics[scale=.8]{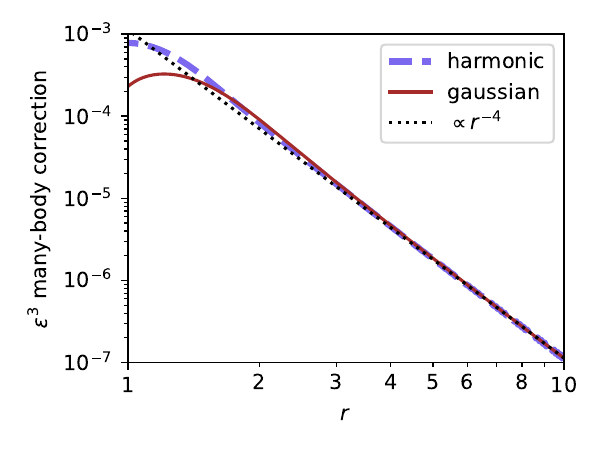}
\end{center}
\caption{Corrections to the correlation that involve the many-body potential $\phi$ at order $\epsilon^3$ for the harmonic potential (thick dashed blue line) and for a Gaussian potential with the same integral (solid red line), with a fit to an algebraic decay as $r^{-4}$.}
\label{fig:comp_harm_gauss}
\end{figure}

We compare the order $\epsilon^3$ of the induced interaction obtained from the harmonic potential and a Gaussian potential with the same integral in Fig.~\ref{fig:comp_harm_gauss}, showing that they share the same prefactor for the algebraic decay. 
This confirms the general form (\ref{eq:gen_pert}).
As the numerical integration is much faster with a Gaussian potential, we use a Gaussian potential to compute the prefactor as a function of the temperature $T_B$ (Fig.~1(b) in the main text).

\begin{figure}
    \begin{center}
           \includegraphics[scale=.6]{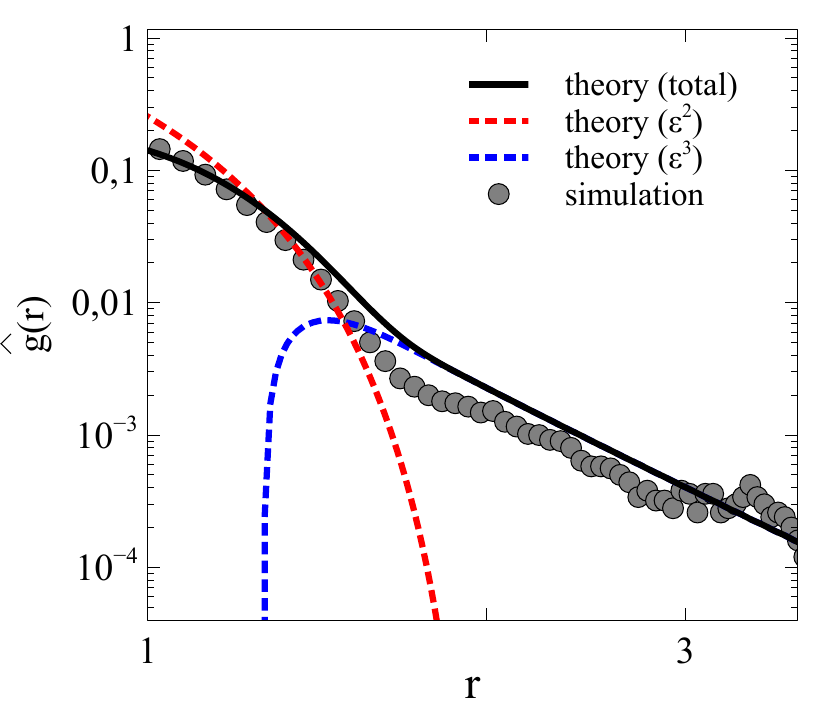}
    \end{center}
    \caption{Correlation $\hat{g}(r)$ for an harmonic interaction, $T_A=1$, $T_B=3$ and $\epsilon=3$: simulations (circles) and theoretical prediction (black solid line) with the contributions of order $\epsilon^2$ and $\epsilon^3$ (red and blue dashed lines, respectively).
    \label{fig:quantitative_comp}}
\end{figure}

The numerical integration also allows a quantitative comparison of the full correlation $\hat g(r)$.
This comparison is presented in Fig.~\ref{fig:quantitative_comp} for an harmonic interaction, $T_A=1$, $T_B=3$ and $\epsilon=3$.

\section{Dependence on the pair potential}

To assess the generality of our results, we ran simulations with other potentials: Gaussian (Eq.~\ref{eq:gauss}), WCA and Lennard-Jones (Eq.~\ref{eq:WCA}).
The results are presented below.

\subsection{Gaussian and WCA repulsive interactions}
\label{sec:sim_gauss_wca}

\begin{figure}[h]
    \begin{center}
      \includegraphics[scale=.4]{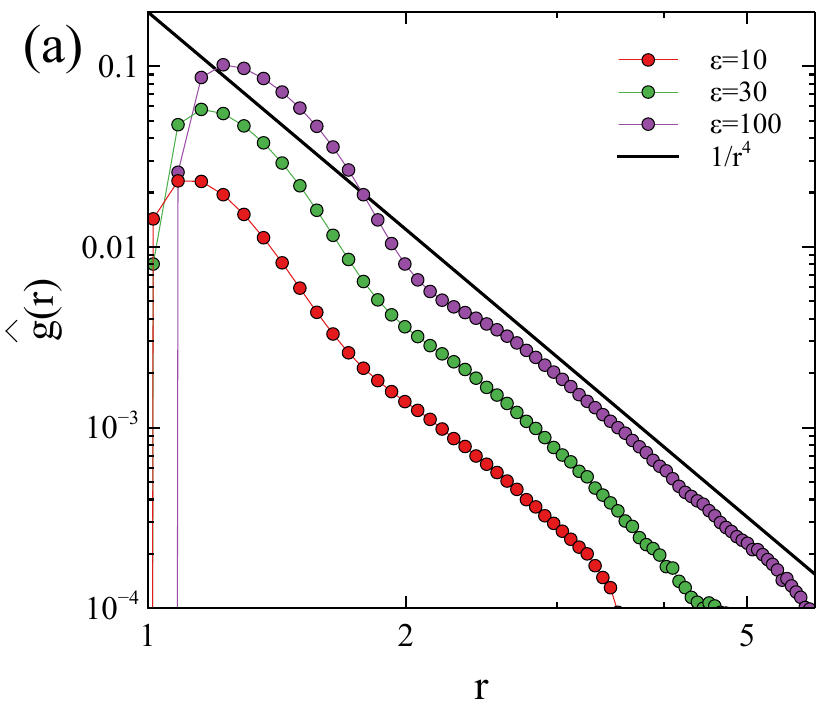}
      \includegraphics[scale=.4]{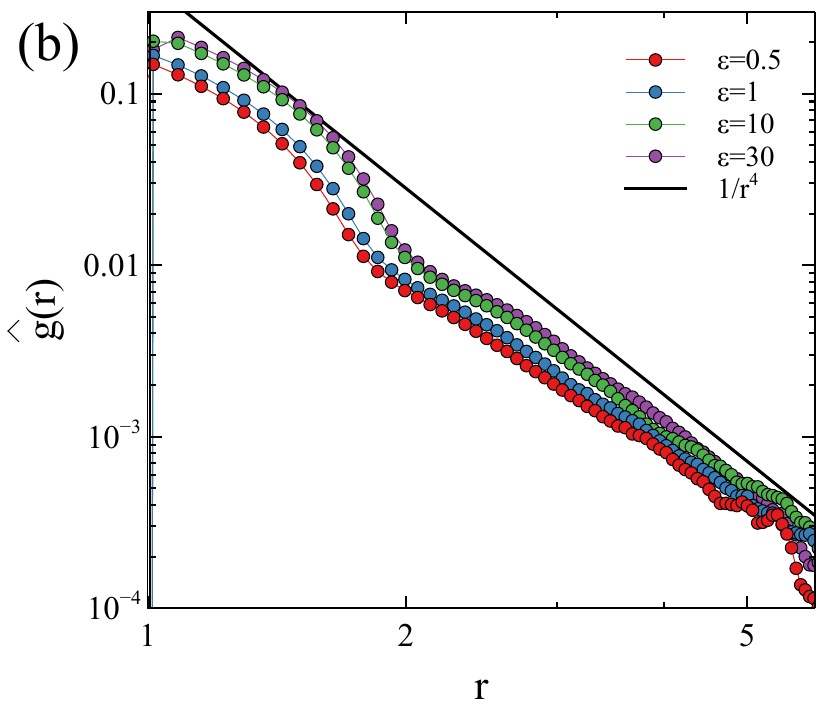}
      \includegraphics[scale=.4]{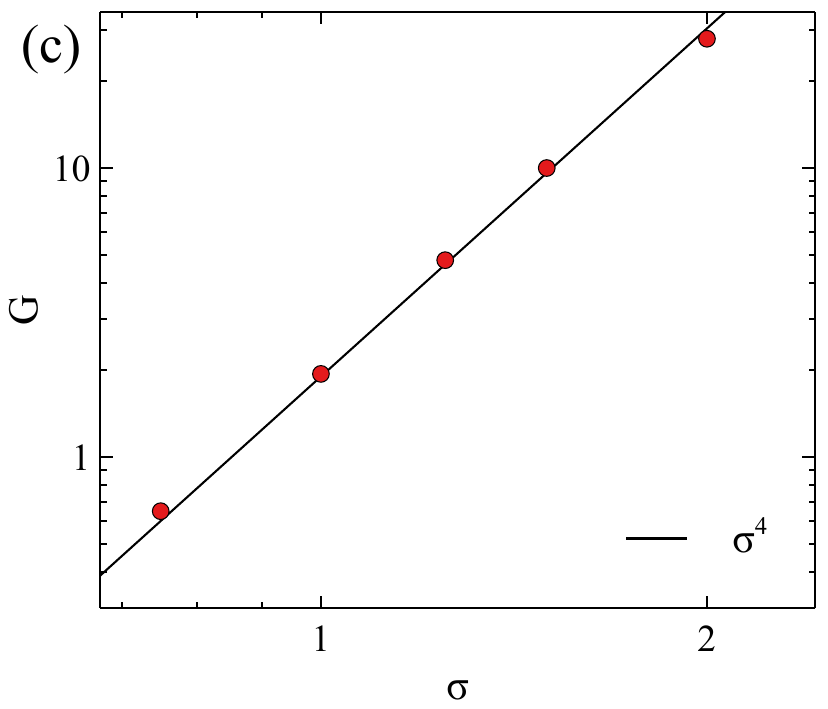}
    \end{center}
    \caption{(a,b) Correlation $\hat{g}(r)$ for a Gaussian (a) and WCA interaction (b), $T_A=1$, $T_B=3$ and various magnitudes of the potential, $\epsilon$ as indicated. 
    (c) Prefactor of the algebraic decay for a hard WCA interaction ($\epsilon=100$), $T_A=1$, $T_B=10$, as a function of the diameter $\sigma$ of the particles (circles). The solid line is the theoretical prediction $G\propto\sigma^4$.}
    \label{fig:size}
\end{figure}

For Gauss and WCA potential, we recover the long range correlations with the same $1/r^4$ dependence, showing the robustness of the theoretical analysis that does not depend on the shape of the potential. 
Besides, the evolution of the prefactor with the diameter $\sigma$ of the particles shown in Fig.~\ref{fig:size} for a hard interaction (WCA with $\epsilon=100$, see Eq.~\ref{eq:WCA}) further confirms the $1/r^4$ scaling.

\subsection{Lennard-Jones interaction}
\label{sec:sim_lj}

Considering a Lennard-Jones potential with attractive and repulsive components is also interesting. 
The correlations computed from simulations exhibit a crossover. 
When $\epsilon>T_B$, the attractive part of the interaction dominates and we observe a complex shape in the correlations characteristic of a structured fluid. In contrast, when $\epsilon<T_B$, the structures observed in the correlations disappear and we recover the $1/r^4$ long range correlation characteristic of purely repulsive potentials (Fig.~\ref{fig:size}). The attractive part of the interactions becomes negligible.

\begin{figure}[h]
    \begin{center}
           \includegraphics[scale=.6]{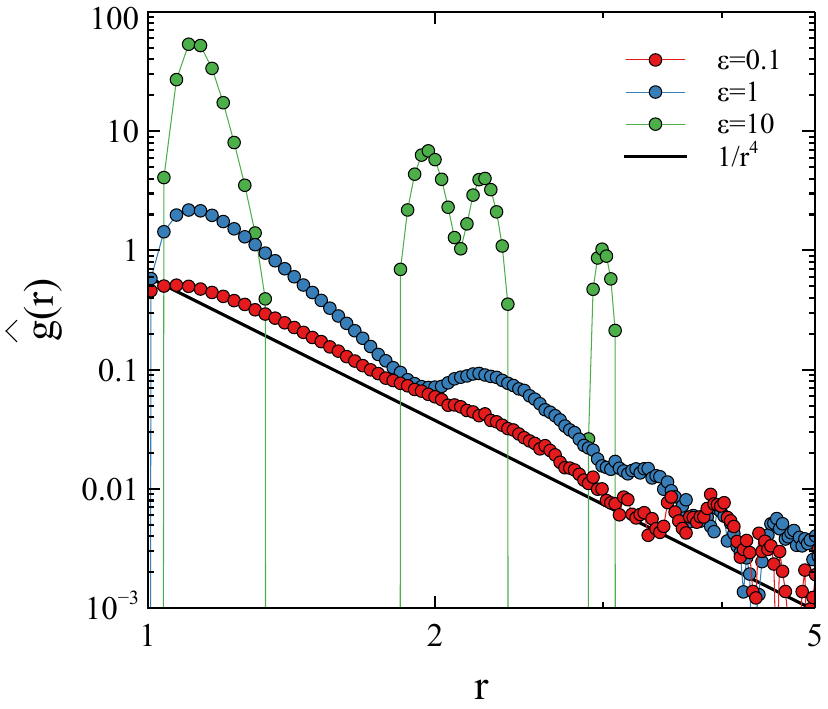}
    \end{center}
    \caption{Correlation $\hat{g}(r)$ for a LJ interaction (i.e., involving an attractive part), $T_A=1$, $T_B=3$ and various magnitudes of the potential, $\epsilon$ as indicated.
    \label{fig:hat_g_LJ}}
\end{figure}


%

\end{document}